\documentclass[twocolumn]{aastex631}

\usepackage{tabularx}
\usepackage{amsmath}  
\usepackage{amssymb}  
\usepackage[T1]{fontenc}
\usepackage{xspace}
\usepackage{verbatim}
\usepackage{natbib}
\defcitealias{pierel_lenswatch}{P23}
\defcitealias{goobar_zwicky}{G23}
\newcommand{\redpen}[1]{{\bf\textcolor{red}{#1}}}

\usepackage{newunicodechar,graphicx}
\DeclareRobustCommand{\okina}{%
  \raisebox{\dimexpr\fontcharht\font`A-\height}{%
    \scalebox{0.8}{`}%
  }%
}
\newunicodechar{ʻ}{\okina}

\newcommand{\angstrom}{\mbox{\normalfont\AA}\xspace}

\def\arcsec{\ensuremath{^{\prime\prime}}}

\def\xone{\ensuremath{x_{1}}}
\def\C{\ensuremath{\mathcal{C}}}

\newcommand{\lensedsn}{SN Zwicky\xspace}

\newcommand{\sntd}{{\fontfamily{qcr}\selectfont{SNTD}}\xspace}
\newcommand{\dolphot}{{\fontfamily{qcr}\selectfont{DOLPHOT}}\xspace}
\newcommand{\gaussn}{{\fontfamily{qcr}\selectfont{GausSN}}\xspace}

\newcommand{\hst}{\textit{HST}\xspace}

\newcommand{\snz}{0.3544}

\def\snstretch{1.16\xspace}
\def\snc{0.005\xspace}

\def\snpeak{59808.6\xspace}
\begin{document}

\title{LensWatch. II. Improved Photometry and Time-delay Constraints on the Strongly Lensed Type Ia Supernova 2022qmx (``SN Zwicky'') with \textit{HST} Template Observations}

\author[0000-0003-2037-4619]{C.~Larison}
\correspondingauthor{Conor~Larison} 
\email{conorjlarison@gmail.com}
\affil{Department of Physics \& Astronomy, Rutgers, State University of New Jersey, 136 Frelinghuysen Road, Piscataway, NJ 08854, USA}

\author[0000-0002-2361-7201
]{J.~D.~R.~Pierel}
\affil{Space Telescope Science Institute, 3700 San Martin Drive, Baltimore, MD 21218, USA}

\author[0000-0002-8092-2077]{M.~J.~B.~Newman}
\affil{Department of Physics \& Astronomy, Rutgers, State University of New Jersey, 136 Frelinghuysen Road, Piscataway, NJ 08854, USA}

\author[0000-0001-8738-6011]{S.~W.~Jha}
\affil{Department of Physics \& Astronomy, Rutgers, State University of New Jersey, 136 Frelinghuysen Road, Piscataway, NJ 08854, USA}

\author[0000-0002-5116-7287]{D.~Gilman}
\affil{Department of Astronomy \& Astrophysics, University of Chicago, Chicago, IL 60637, USA}

\author[0000-0003-3847-0780]{E.~E.~Hayes}
\affil{Institute of Astronomy and Kavli Institute for Cosmology, University of Cambridge, Madingley Road, Cambridge CB3 0HA, UK}

\author[0009-0008-1965-9012]{A.~Agrawal}
\affil{Department of Astronomy, University of Illinois at Urbana-Champaign, 1002 W. Green St., IL 61801, USA}

\author[0000-0001-5409-6480]{N.~Arendse}
\affil{Oskar Klein Centre, Department of Physics, Stockholm University, SE-10691 Stockholm, Sweden}

\author[0000-0003-3195-5507]{S.~Birrer}
\affil{Department of Physics and Astronomy, Stony Brook University, Stony Brook, NY 11794, USA}

\author[0000-0002-1537-6911]{M.~Bronikowski}
\affil{Center for Astrophysics and Cosmology, University of Nova Gorica, Vipavska 11c, 5270 Ajdov\v{s}\v{c}ina, Slovenia}

\author[0000-0001-6711-8140]{S.~Chakrabarti}
\affil{Department of Physics and Astronomy, University of Alabama, Huntsville, Huntsville, Alabama 35899}

\author[0000-0003-0928-2000]{J.~M.~Della~Costa}
\affil{NSF NOIRLab, Tucson, AZ 85719, USA}

\author[0000-0003-4263-2228]{D.~A.~Coulter}
\affil{Space Telescope Science Institute, 3700 San Martin Drive, Baltimore, MD 21218, USA}

\author[0000-0003-0758-6510]{F.~Courbin}
\affil{ICC-UB Institut de Ci\`encies del Cosmos, Universitat de Barcelona, Mart\'i Franqu\`es, 1, E-08028 Barcelona, Spain}
\affil{ICREA, Pg. Llu\'is Companys 23, Barcelona, E-08010, Spain}

\author[0000-0001-8737-9700]{K.~A.~Dalrymple}
\affil{Johns Hopkins University, William H. Miller III Department of Physics and Astronomy, 3400 North Charles St., Baltimore, MD 21218, USA}

\author[0000-0002-2376-6979]{S.~Dhawan}
\affil{Institute of Astronomy and Kavli Institute for Cosmology, University of Cambridge, Madingley Road, Cambridge CB3 0HA, UK}

\author[0000-0001-9065-3926]{J.~M.~Diego}
\affil{Instituto de F\'isica de Cantabria (CSIC-UC), Avda. Los Castros s/n, 39005 Santander, Spain}

\author[0000-0002-8526-3963]{C.~Gall}
\affil{DARK, Niels Bohr Institute, University of Copenhagen, Jagtvej 155A, 2200 Copenhagen, Denmark}

\author[0000-0002-4163-4996]{A.~Goobar}
\affil{Oskar Klein Centre, Department of Physics, Stockholm University, SE-10691 Stockholm, Sweden}

\author[0000-0002-4571-2306]{J. Hjorth}
\affil{DARK, Niels Bohr Institute, University of Copenhagen, Jagtvej 155A, 2200 Copenhagen, Denmark}

\author[0000-0001-8156-0330]{X.~Huang}
\affil{Department of Physics \& Astronomy, University of San Francisco, San Francisco, CA 94117}

\author[0000-0001-5975-290X]{J.~Johansson}
\affil{Oskar Klein Centre, Department of Physics, Stockholm University, SE-10691 Stockholm, Sweden}

\author[0000-0001-8317-2788]{S.~Mao}
\affil{Department of Astronomy, Tsinghua University, Beijing 100084 China}

\author[0000-0001-8442-1846]{R.~Marques-Chaves}
\affil{Department of Astronomy, University of Geneva, 51 Chemin Pegasi, 1290 Versoix, Switzerland}

\author[0000-0001-6876-8284]{P.~A.~Mazzali}
\affil{Astrophysics Research Institute, Liverpool John Moores University, IC2, Liverpool Science Park, 146 Brownlow Hill, Liverpool L3 5RF, UK}
\affil{Max-Planck-Institut für Astrophysik, Karl-Schwarzschild Straße 1, 85748 Garching, Germany}

\author[0000-0001-7714-7076]{A.~More}
\affil{Inter-University Centre for Astronomy and Astrophysics, Ganeshkhind, Pune 411007, India}
\affil{Kavli Institute for the Physics and Mathematics of the Universe (WPI), University of Tokyo, 5-1-5, Kashiwa, Chiba 277-8583, Japan}

\author[0000-0003-3030-2360]{L.~A.~Moustakas}
\affil{Jet Propulsion Laboratory, California Institute of Technology}

\author[0000-0002-2807-6459]{I. P\'{e}rez-Fournon}
\affil{Instituto de Astrof\'\i sica de Canarias, C/V\'\i a L\'actea, s/n, E-38205 San Crist\'obal de La Laguna, Tenerife, Spain}
\affil{Universidad de La Laguna, Dpto. Astrof\'\i sica, E-38206 San Crist\'obal de La Laguna, Tenerife, Spain}

\author[0000-0003-4743-1679]{T.~Petrushevska}
\affil{Center for Astrophysics and Cosmology, University of Nova Gorica, Vipavska 11c, 5270 Ajdov\v{s}\v{c}ina, Slovenia.}

\author[0000-0002-5391-5568]{F.~Poidevin}
\affil{Instituto de Astrof\'{\i}sica de Canarias, V\'{\i}a   L\'actea, 38205 La Laguna, Tenerife, Spain}
\affil{Universidad de La Laguna, Departamento de Astrof\'{\i}sica,  38206 La Laguna, Tenerife, Spain}

\author[0000-0002-4410-5387]{A.~Rest}
\affil{Johns Hopkins University, William H. Miller III Department of Physics and Astronomy, 3400 North Charles St., Baltimore, MD 21218, USA}
\affil{Space Telescope Science Institute, 3700 San Martin Drive, Baltimore, MD 21218, USA}

\author[0000-0002-5558-888X]{A.~J.~Shajib}
\affil{Department of Astronomy \& Astrophysics, University of Chicago, Chicago, IL 60637, USA}
\affil{Kavli Institute for Cosmological Physics, University of Chicago, Chicago, IL 60637, USA}

\author[0000-0002-1114-0135]{R.~Shirley}
\affil{Max-Planck-Institut für extraterrestrische Physik, Giessenbachstr. 1, 85748 Garching, Germany}

\author[0000-0002-7756-4440]{L.~G.~Strolger}
\affil{Space Telescope Science Institute, 3700 San Martin Drive, Baltimore, MD 21218, USA}

\author[0000-0001-5568-6052]{S.~H.~Suyu}
\affil{Technical University of Munich, TUM School of Natural Sciences, Physics Department, James-Franck-Str.~1, 85748 Garching, Germany}
\affil{Max-Planck-Institut für Astrophysik, Karl-Schwarzschild Straße 1, 85748 Garching, Germany}

\author[0000-0002-8460-0390]{T.~Treu}
\affil{Physics and Astronomy Department University of California Los Angeles CA 90095}

\author[0000-0002-0632-8897]{Y.~Zenati}
\affil{Johns Hopkins University, William H. Miller III Department of Physics and Astronomy, 3400 North Charles St., Baltimore, MD 21218, USA}
\affil{Space Telescope Science Institute, 3700 San Martin Drive, Baltimore, MD 21218, USA}

\begin{abstract}
Strongly lensed supernovae (SNe) are a rare class of transient that can offer tight cosmological constraints that are complementary to methods from other astronomical events. We present a follow-up study of one recently-discovered strongly lensed SN, the quadruply-imaged Type Ia SN 2022qmx (aka, ``SN Zwicky'') at $z=\snz$. We measure updated, template-subtracted photometry for \lensedsn and derive improved time delays and magnifications. This is possible because SNe are transient, fading away after reaching their peak brightness. Specifically, we measure point spread function (PSF) photometry for all four images of \lensedsn in three \emph{Hubble Space Telescope} WFC3/UVIS passbands (F475W, F625W, F814W) and one WFC3/IR passband (F160W), with template images taken $\sim 11$ months after the epoch in which the SN images appear. We find consistency to within $2\sigma$ between lens model predicted time delays ($\lesssim1$ day), and measured time delays with \hst colors ($\lesssim2$ days), including the uncertainty from chromatic microlensing that may arise from stars in the lensing galaxy. The standardizable nature of SNe Ia allows us to estimate absolute magnifications for the four images, with images A and C being elevated in magnification compared to lens model predictions by about $6\sigma$ and $3\sigma$ respectively, confirming previous work. We show that millilensing or differential dust extinction is unable to explain these discrepancies and find evidence for the existence of microlensing in images A, C, and potentially D, that may contribute to the anomalous magnification.
\end{abstract}

\section{Introduction}
\label{sec:intro}
Supernovae (SNe) that have been multiply-imaged by strong gravitational lensing are naturally rare and revealing astronomical events. They require an unlikely alignment along the line of sight between an observer, the background source that is lensed, and the foreground lens (galaxy or cluster). This, combined with the fact that most SNe rise and fade over the course of a few rest-frame weeks, makes strongly lensed SNe extremely elusive. 

The geometry of the lensing system and gravitational potential differences across the lens plane determine the delay in arrival of the SN images relative to one another. Measurements of this ``time delay'' can provide an angular diameter distance, which in turn can constrain the Hubble constant ($H_0$) and the dark energy equation of state ($w$) directly  \citep[e.g.,][]{refsdal_possibility_1964,linder_lensing_2011,paraficz_gravitational_2009,treu_time_2016,treu_strong_2022,birrer_time-delay_2022}. Strongly lensed SNe have several advantages over strongly lensed quasars, which have often been used for time-delay measurements \citep[e.g.,][]{vuissoz_cosmograil_2008,suyu_dissecting_2010,tewes_cosmograil_2013,bonvin_h0licow_2017,birrer_h0licow_2019,bonvin_cosmograil_2018,bonvin_cosmograil_2019,wong_h0licow_2020}:

\begin{itemize}
    \item SNe fade on a short timescale (over weeks to months), so more accurate models of the lensing system can be made post-SN discovery, as the source and lens fluxes would remain highly blended otherwise \citep{ding_improved_2021}.
    \item SNe (especially of Type Ia) have predictable light curves, thus simplifying time-delay measurements over a more stochastic system such as an active galactic nucleus (AGN).
    \item Strongly lensed SNe are also affected by microlensing \citep{dobler_micro_2006,tie_microlensing_2018,foxley-marrable_impact_2018,bonvin_impact_2019}; however, chromatic effects are mitigated given sufficient early-time light curve coverage  \citep{goldstein_precise_2018,foxley-marrable_impact_2018,huber_strongly_2019}. Microlensing can still be a significant source of uncertainty for SNe with small time delays \citep[e.g.,][]{goobar_iptf16geu_2017,more_interpreting_2017} as we will show later.
    \item Perhaps most important in terms of logistical constraints is that lensed SNe require much shorter observing campaigns than lensed AGN, potentially going from a decade of extended observations for a single lensed AGN down to a few epochs over a year for a strongly lensed SN \citep{pascale_h0pe,pierel_h0pe}.
\end{itemize}

These advantages been shown and explored for decades \citep[e.g.,][]{refsdal_possibility_1964,kelly_multiple_2015,goobar_iptf16geu_2017, goldstein_precise_2018,huber_strongly_2019,pierel_turning_2019,suyu_holismokes_2020,pierel_projected_2021,rodney_gravitationally_2021,goobar_lessons_2024}, but observations to date have not been optimized for this type of phenomenon, as evidenced by only a few detections being made by recent wide-field astronomical surveys \citep{goobar_iptf16geu_2017,goobar_zwicky,townsend_glsnia_2024}. Despite few discoveries, we now have a confirmed sample of eight multiply-imaged SNe (SN Refsdal, SN 2016geu, SN Requiem, C22, SN Zwicky, SN 2022riv, SN H0pe, and SN Encore), including two that are lensed by a galaxy \citep{goobar_iptf16geu_2017,goobar_zwicky,pierel_lenswatch} and six that have been lensed by a foreground galaxy cluster \citep{kelly_multiple_2015,rodney_gravitationally_2021,chen_cooling_2022,kelly_strongly_2022,frye_h0pe,pierel_lensed_2024}. Our expanding sample size is a direct result of the combined efforts of many dedicated programs to find these rare events \citep[e.g.,][]{petrushevska_high-redshift_2016,petrushevska_searching_2018,fremling_zwicky_2020,craig_targeted_2021}. 

Type Ia SNe (SNe\,Ia) are especially useful as strongly lensed sources as we possess well-constrained templates of their light curves \citep{hsiao_k_2007,guy_supernova_2010,saunders_snemo_2018,leget_sugar_2020,kenworthy_salt3_2021,pierel_salt3-nir_2022}, which can be used to standardize their brightness \citep{phillips_absolute_1993,tripp_two-parameter_1998}. This standardizability can be used to break the mass-sheet degeneracy, a large systematic effect where one could add additional sheets of mass to the lensing plane without influencing the image positions and flux ratios \citep{falco_model-dependent_1985,kolatt_gravitational_1998,holz_seeing_2001,oguri_gravitational_2003,patel_three_2014,nordin_lensed_2014,rodney_illuminating_2015,xu_lens_2016,birrer_hubble_2022}, albeit only when millilensing and microlensing are mostly mitigated \citep[see][]{goobar_iptf16geu_2017,foxley-marrable_impact_2018,dhawan_magnification_2019,weisenbach_micro_2024}. 

We have been fortunate enough to discover three cluster-scale SN lensing systems with time delays that can precisely measure $H_0$: SN Refsdal, SN H0pe and SN Encore \citep{kelly_refsdal_h0,kelly_refsdal_images,chen_h0pe,frye_h0pe,pascale_h0pe,pierel_h0pe,pierel_lensed_2024,dhawan_encore_2024}. In contrast, the two existing galaxy-scale systems with lensed SNe do not currently provide such precise cosmological results, but they remain important to investigate, as they are expected to be much more common than these cluster-scale lenses in the era of the Vera C. Rubin Observatory \citep{quimby_detection_2014,collett_population_2015,goldstein_rates_2019,wojtak_magnified_2019,sainz_de_murieta_rates,arendse_detecting_2024}. With a larger statistical sample of such systems, it then becomes possible to measure $H_0$ to a percent level, enough to make it competitive in the current field \citep{huber_strongly_2019, suyu_holismokes_2020, birrer_h0_2024, suyu_lensedsne_2024}. However, in order to achieve this level of precision, we must first measure the time delays of each of our SN systems in the sample to a similarly high precision. To do this, we must better understand the impacts of microlensing to our error budget, the shortfalls of our current lens models on galaxy-galaxy systems, and the importance of follow-up campaigns to increase the quality and quantity of our photometry.

In this paper, we use additional observations to investigate one such galaxy-scale system, \lensedsn. \lensedsn was discovered in 2022 August by the Zwicky Transient Facility \citep[ZTF;][]{fremling_zwicky_2020}\footnote{\url{https://www.wis-tns.org/object/2022qmx}}, and was subsequently classified and analyzed by \citet[][hereafter G23]{goobar_zwicky}. The time delays, magnifications, and lens models of the SN \& galaxy-galaxy system were then analyzed and reported in \citet[][hereafter P23]{pierel_lenswatch}. This work represents the second paper in a series of papers for the LensWatch program\footnote{\url{https://www.lenswatch.org}}, a direct follow-up to \citetalias{pierel_lenswatch} that provides improved time-delay and magnification measurements of \lensedsn from template-subtracted \hst photometry. Section \ref{sec:obs} presents the template \hst observation characteristics of \lensedsn. Our analysis of \lensedsn (including photometry and measurements of time delays and magnifications, as well as analysis investigating microlensing and millilensing) are reported in Section \ref{sec:zwicky_analysis}. Finally, we conclude with a discussion of the implications of these results in Section \ref{sec:conclusion}.

\section{LensWatch Observations of \lensedsn with \textit{HST}}
\label{sec:obs}

\subsection{LensWatch Observations Post-discovery}
\label{sub:zwicky}

As \citetalias{pierel_lenswatch} summarizes, roughly $12$ days after the spectroscopic classification of \lensedsn, we used a non-disruptive \hst target of opportunity (ToO) trigger to obtain WFC3/UVIS and IR images of the lensing system. Observations of \lensedsn were made with WFC3/UVIS ($0.04\arcsec$/pix) to resolve the multiple images, specifically in the F475W, F625W, and F814W filters to provide non-overlapping coverage across the full optical wavelength range ($\sim$ 3,500--6,000~\angstrom in the rest-frame). Additionally, we included WFC3/IR F160W observations to provide overall calibration to ground-based NIR data and potentially useful information about the lensing system.

\begin{table*}
    \centering
    \caption{\label{tab:im_mags}Photometry and colors measured for each image of \lensedsn in AB magnitudes from \citetalias{pierel_lenswatch} (top) and our new results from template subtraction.}
    \begin{tabular*}{\linewidth}{@{\extracolsep{\stretch{1}}}*{7}{c}}
\toprule
Image&\multicolumn{1}{c}{F475W}&\multicolumn{1}{c}{F625W}&
\multicolumn{1}{c}{F814W}&\multicolumn{1}{c}{F160W}\\
\hline
A (\citetalias{pierel_lenswatch})&23.22$\pm$0.04&21.67$\pm$0.02&20.67$\pm$0.01&$--$\\
A (Updated)&23.21$\pm$0.03&21.67$\pm$0.02&20.56$\pm$0.01&21.79$\pm$0.01\\
\hline
B (\citetalias{pierel_lenswatch})&24.31$\pm$0.07&22.65$\pm$0.03&21.71$\pm$0.02&$--$\\
B (Updated)&24.13$\pm$0.04&22.48$\pm$0.02&21.45$\pm$0.02&22.35$\pm$0.02\\
\hline
C (\citetalias{pierel_lenswatch})&23.35$\pm$0.04&21.90$\pm$0.02&20.88$\pm$0.02&$--$\\
C (Updated)&23.25$\pm$0.03&21.72$\pm$0.02&20.70$\pm$0.01&21.95$\pm$0.01\\
\hline
D (\citetalias{pierel_lenswatch})&24.26$\pm$0.07&22.72$\pm$0.04&21.60$\pm$0.02&$--$\\
D (Updated)&24.05$\pm$0.04&22.42$\pm$0.02&21.39$\pm$0.02&22.85$\pm$0.02\\
\hline
    \end{tabular*}

\tablenotetext{}{We note that the uncertainties reported by \citetalias{pierel_lenswatch} are underestimated, as systematic uncertainties due to the variable background from the lens galaxy were not taken into account.}

\end{table*}

\subsection{Template Observations}
\label{sub:templates}

As part of the \hst Cycle 28 LensWatch program, we are able to take template observations of lensed SN targets after the SN light has faded. As mentioned in Section\,\ref{sec:intro}, one of the main advantages of studying strongly lensed SNe is that they are transient events that almost completely fade on the order of months (depending on the SN type and observed filter). Template observations were taken August 10, 2023, approximately 11 months after the LensWatch ToO \hst observations of \lensedsn. This corresponds to approximately 8 months in the supernova rest-frame, when a typical SN~Ia has an optical luminosity less than about 1\% of its peak \citep{jha_observational_2019}, a negligible contribution for our measurements.

Templates were obtained in all four filters used and summarized in \citetalias{pierel_lenswatch}: WFC3/UVIS F475W, F625W, F814W, and WFC3/IR F160W. To avoid any unnecessary issues with aligning and scaling SN image and template observations, we used the same dithering patterns, pointings, position angles, and exposure times as were used in the initial observations. Each filter had three dither positions, resulting in a total of 12 exposures across all filters.

\section{Analysis and Results}
\label{sec:zwicky_analysis}
\subsection{\hst Template Photometry}
\label{sub:phot}

For future discovered strongly lensed transients, it may not always be possible to obtain template images of every strongly lensed SN in all observed filters, although workarounds are being investigated, including using data from \textit{Euclid} (Akhaury et al., in prep.). Therefore, it is important to examine how well we can measure image fluxes with and without the template photometry we have available.

\begin{figure*}
    \centering
    \includegraphics[scale=0.9]{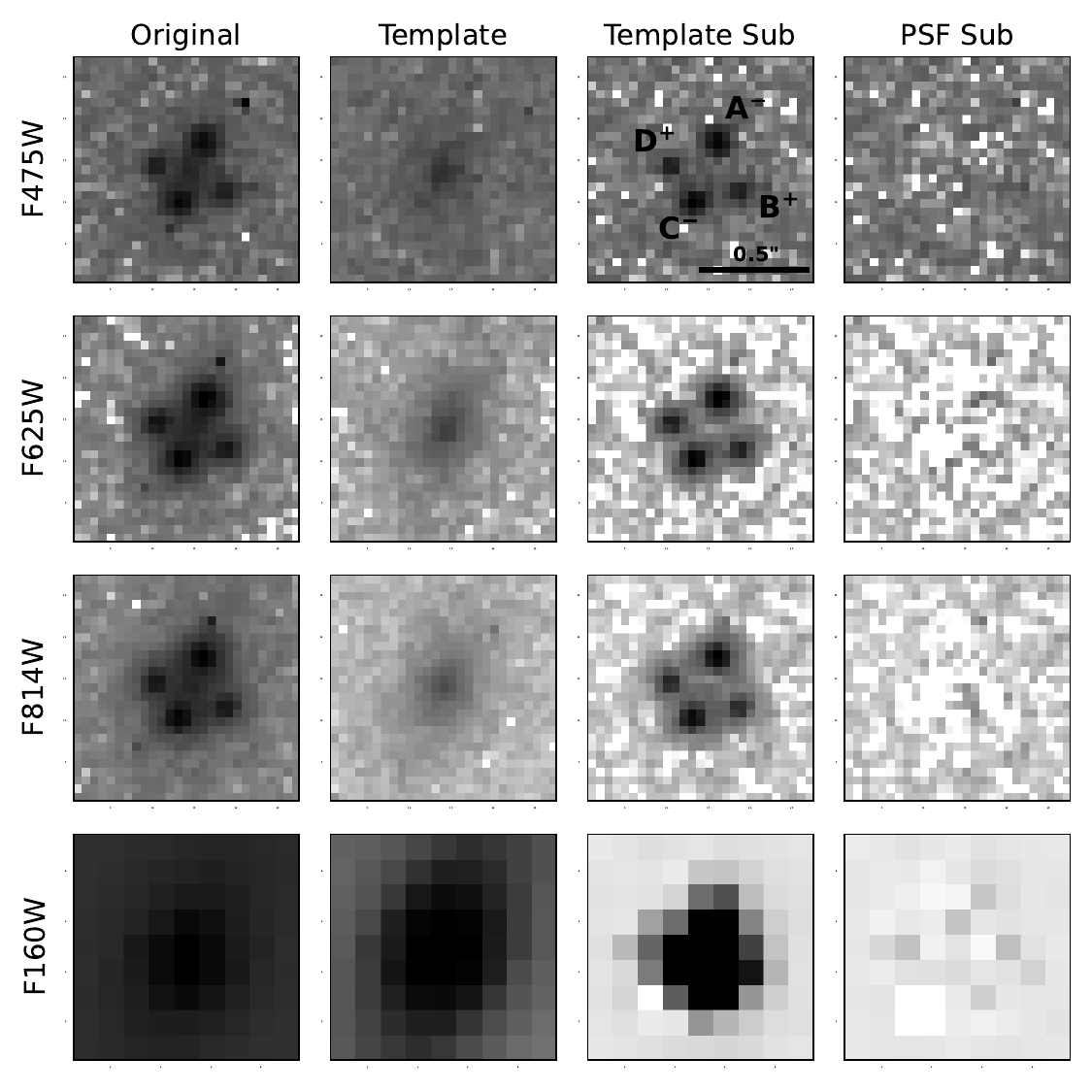}
    \caption{Images of our original HST observations, template observations, template-subtracted images, and PSF-subtracted images for each of our WFC3/UVIS filters, each from a single FLC. The images are reprojected, such that up is celestial North and left is East. Final fluxes were calculated in parallel among all FLC images. We also include the same set of images but for a WFC3/IR F160W FLT frame, with the PSF-subtracted images from the \dolphot routine described in Section~\ref{sub:phot}. The SN images, A-D, are labelled in the third panel of row one, with the magnification parities labelled as a superscript above each image letter.}
    \label{fig:psf_sub}
\end{figure*}

In order to do PSF photometry, we first subtract our template images from the exposures containing the SN images. Due to careful planning of our follow-up template epochs, we were able to align the images using a simple pixel offset without a need to align to a stellar or SDSS object catalog for the UVIS filters (we will discuss the alignment process for the IR filter later, as it differed in scope). We were also able to make subtractions without scaling our observations due to the stability of the HST instruments and our consistent exposure times.

Our subtractions were done using the SN and template WFC3/UVIS ``FLC'' images, which are individual exposures that have been bias-subtracted, dark-subtracted, and flat-fielded but not yet corrected for geometric distortion. Because the PSF solution can vary across the detector, it is important to use these images that are still uncorrected for distortion for PSF photometry instead of final ``drizzled'' products, which can introduce inconsistencies into the modeling of a PSF. We use the standard \hst PSF models\footnote{\url{https://www.stsci.edu/hst/instrumentation/wfc3/data-analysis/psf}} to represent the PSF, which also take into account spatial variation across the detector.
For our WFC3/IR subtractions, we used the ``FLT'' image, which lacks a charge transfer efficiency (CTE) correction that is necessary for UVIS images, but is not applicable to the IR data.

For each UVIS filter, we use centroiding to determine the positions of all four SN images across the three FLCs. We also estimate a uniform background for each image position, using a mode estimator algorithm. For each UVIS filter, we then perform forced PSF photometry by implementing a Bayesian nested sampling routine\footnote{\textsc{dynesty}: \url{https://dynesty.readthedocs.io/en/stable/}} to constrain the (common) SN flux in all three FLCs for all four SN images using the package, \textsc{space\_phot}\footnote{\url{https://space-phot.readthedocs.io/en/latest/}} \citep{pierel_space_phot_2024}. Each PSF was fit to the multiple SN images within a $5\times5$ pixel square to limit the contamination from the other SN images, which captures $\sim99\%$ of the total SN flux given the FWHM of $\sim 2$ pixels.

\begin{table*}[!t]
    \centering
    \caption{\label{tab:td_mu}Updated time delays and magnifications (with parities from lens models) compared to the predictions from lens models and measurements of \citetalias{pierel_lenswatch}}
    \begin{tabular*}{\linewidth}{@{\extracolsep{\stretch{1}}}*{8}{c}}

\toprule
Image&$(\Delta t_{iA})_{\rm{updated}}$&$(\Delta t_{iA})_{\rm{P23~meas}}$&$(\Delta t_{iA})_{\rm{P23~pred}}$&$|\mu_{\rm{updated}}|$&$|\mu_{\rm{P23~meas}}|$&$|\mu_{\rm{P23~pred}}|$\\
&Days&Days&&\\
\hline
A ($-$) &--&--&--&$\redpen{9.13^{+5.21}_{-0.85}}$&$8.31^{+4.16}_{-1.43}$
&$1.81^{+0.90}_{-0.89}$\\
B ($+$)&$\redpen{0.52^{+2.11}_{-1.55}}$&$0.30^{+3.51}_{-3.22}$&$-0.50^{+0.15}_{-0.21}$&$\redpen{4.04^{+1.85}_{-0.50}}$&$3.24^{+1.69}_{-0.57}$
&$3.72^{+1.04}_{-1.24}$\\
C ($-$)&$\redpen{1.97^{+2.28}_{-1.50}}$&$0.30^{+3.40}_{-3.59}$&$-0.22^{+0.10}_{-0.10}$&$\redpen{7.78^{+3.88}_{-0.81}}$&$6.73^{+3.38}_{-1.16}$
&$2.87^{+1.51}_{-1.50}$\\
D ($+$)&$\redpen{0.78^{+2.03}_{-1.69}}$&$0.19^{+3.53}_{-2.97}$&$-0.42^{+0.12}_{-0.18}$&$\redpen{4.22^{+1.99}_{-0.46}}$&$3.39^{+1.65}_{-0.62}$
&$4.12^{+1.19}_{-1.36}$\\
\hline
    \end{tabular*}
    
\end{table*}

The final measured flux is the integral of each PSF model; then, corrected fluxes were converted to AB magnitudes using the time-dependent inverse sensitivity and filter pivot wavelengths provided with each data file. The final measured magnitudes and colors are reported in Table \ref{tab:im_mags} and compared to the same measurements obtained in \citetalias{pierel_lenswatch}. We note that the uncertainties on the values from \citetalias{pierel_lenswatch} are underrepresented, as systematic uncertainties due to the variable background from the lens galaxy were not taken into account. Therefore, the differences seen between these and the updated measurements are less statistically significant than they may first appear. The updated fluxes are larger than the previous measuremeents due to an oversubtraction of the background for the latter, which resulted in large residuals at the image positions, as seen in Figure~7 of \citetalias{pierel_lenswatch}.

We also perform PSF photometry for our WFC3/IR F160W subtractions. We align each original FLT image to its template image using \textsc{Astroalign} and then create subtractions \citep{beroiz_astroalign}. Our fitting method with \textsc{space\_phot} does not handle blended sources, as the PSFs are all fit and subtracted individually, so we turn to a new method using \dolphot \citep{dolphot}. 

\dolphot was created to provide accurate PSF photometry for stellar sources in crowded fields in \hst images. Unlike the PSF fitting routine described above, \dolphot provides iterative PSF photometry, fitting the brighter PSF sources and then subtracting them before fitting fainter nearby sources. We attempt thus to apply \dolphot to our F160W data. First, we take the inner portion of our subtracted FLT images and replace the centers of our template images with these subtracted ``stamps'', so that the sky gradient remains roughly continuous for \dolphot to accurately measure the sky background. Then, we pedestal that subtracted region to match the background of the template image, so that the sky gradient on the full image is smooth.

In order to run \dolphot on the F160W frames, in which the SN images are not very well separated, we rely on accurate positions from the original F475W FLT images that have clearly resolved SN images instead. To do this, we do a first-pass at alignment between the F475W drizzled image and the F160W drizzled image using \textsc{TweakReg} \citep{fruchter_drizzlepac}. We then run \dolphot on all the F475W frames and the F160W frames simultaneously, with the F475W drizzled image as the reference image for alignment. The PSF photometry obtained for the F475W images agrees with what was obtained by \citetalias{pierel_lenswatch} to within 1$\sigma$, thus providing a check on the accuracy of the method compared to the method outlined above. Adding the measured PSF fluxes from \dolphot and using aperture photometry on the full, unresolved flux from the four SN images in the F160W band, we find that the measured \dolphot fluxes account for $\sim$98\% of the total observed flux. 

We show the results of our PSF modelling and subtraction, as well as the template subtractions in Figure\,\ref{fig:psf_sub}.

\begin{figure}
    \centering
    \includegraphics[scale=0.57]{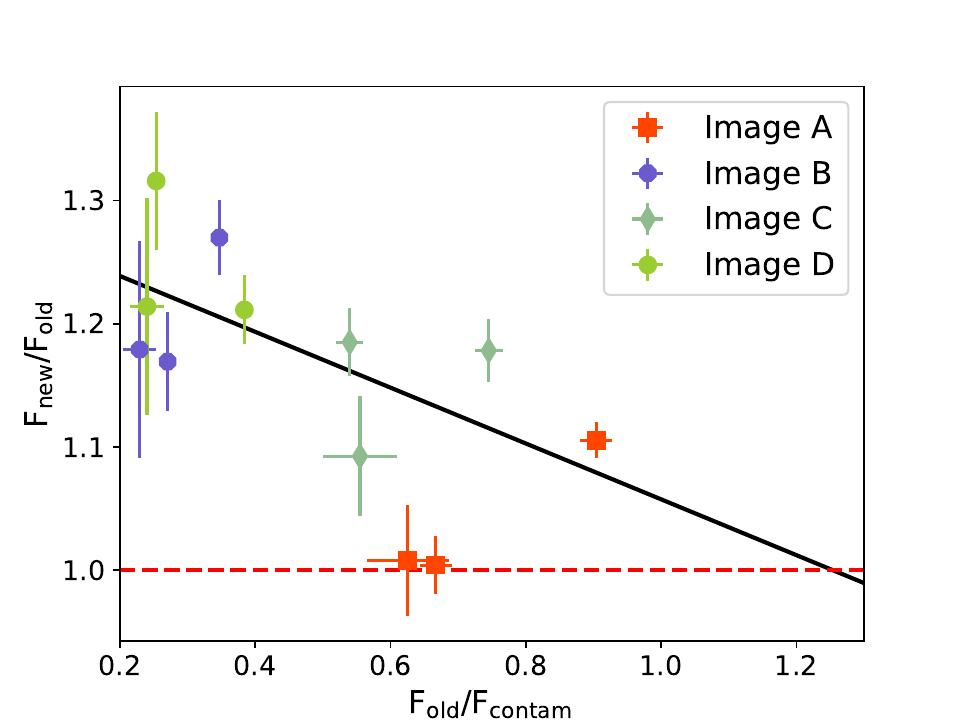}
    \caption{Ratio of new, template photometry flux to old, contaminated flux on y-axis and ratio of old SN flux to lens (and host) galaxy flux from template images on x-axis for each SN image. When the SN flux is around 1.2 times brighter than the contaminating lens and host galaxy flux, the change in SN photometry appears to be negligible, as shown by the linear fit. We combine data from all filters. }
    \label{fig:flux_trends}
\end{figure}

We postulate that the brighter the SN image is when compared to the total flux of the contaminating lens + host galaxy, the less the photometry will change with a template image. Comparing the template-subtracted photometry to the original, we expect the brighter SN images to be affected less by the background light. In Figure~\ref{fig:flux_trends} we test this by examining the ratio of the image fluxes after and before template-subtraction as a function of the ratio of the old flux to the contaminating flux. To measure the contaminating flux, we perform aperture photometry using \textsc{space\_phot} on the ``drizzled'' data products of our template observations, centered on the lensing galaxy. This code accounts for aperture corrections as well as sky background and propagates these effects into a final error that we include in this analysis. We find that the ratio of the old flux to the lens flux for the flux change to be negligible is $1.25 \pm 0.59$. Therefore, if the flux of the SN is about equal to, or greater than, the contaminating flux of the lens + host galaxies, the photometry will probably be robust to changes with template imaging. In the case of \lensedsn, we do not have photometry past this critical point, so we can not test if the bias from the contaminating flux persists, but this would be an important test for future galaxy-scale lens systems.

\begin{figure*}
    \centering
    \includegraphics[width=\linewidth]{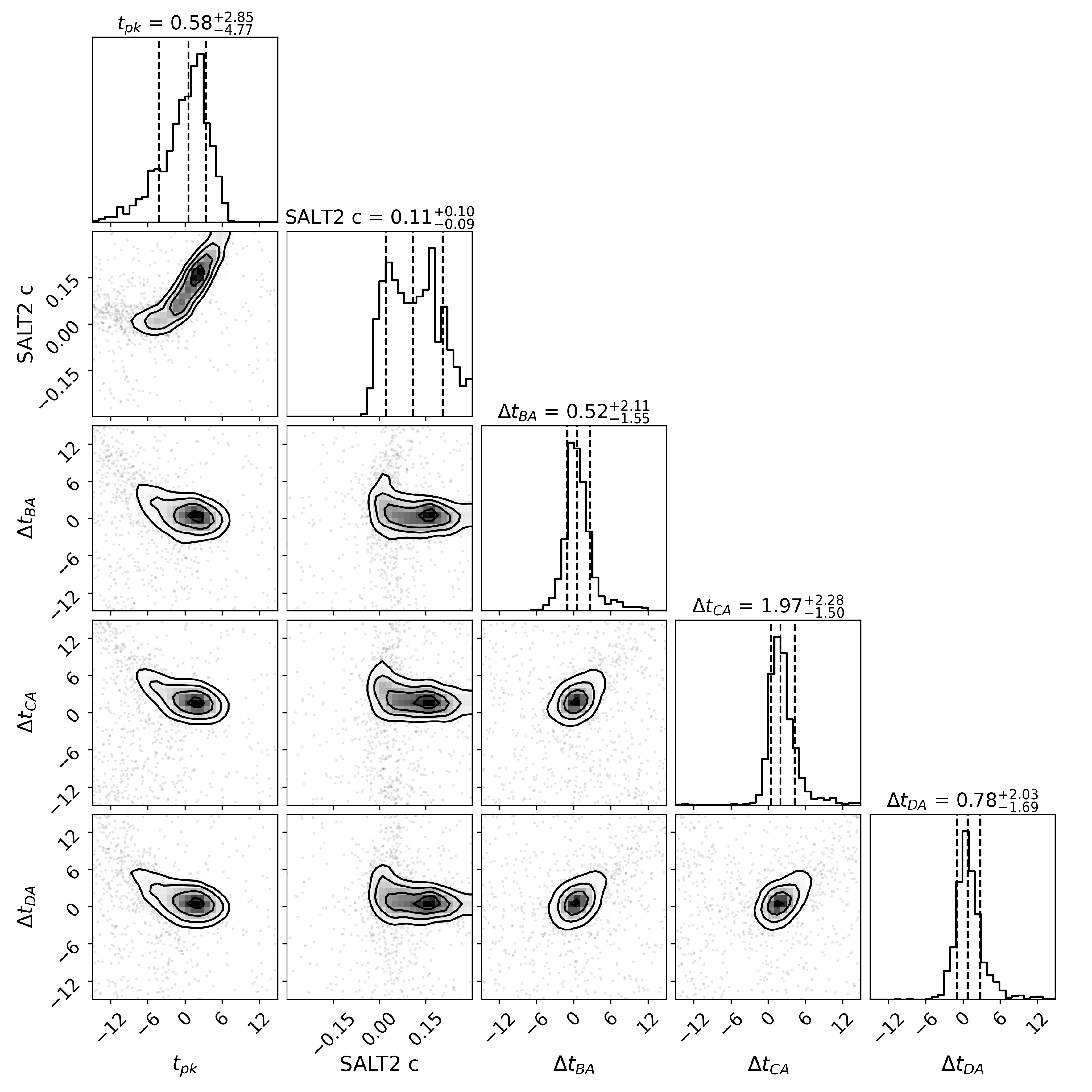}
    \caption{Posterior distributions of $t_{pk}$, the SALT2 color parameter, $c$, and the time delays (relative to image A) from the \sntd Color method fitting to the WFC3/UVIS template photometry. The dashed vertical lines correspond to the distribution $16^{\rm{th}} , \ 50^{\rm{th}}, \ \rm{and} \ 84^{\rm{th}}$ weighted quantiles. The contours on the 2D posteriors correspond to the (0.5, 1.0, 1.5, 2.0) $\sigma$ level contours. The $t_{pk}$ parameter is given relative the MJD date: \snpeak, from the unresolved lightcurve fit in \citetalias{goobar_zwicky}.}
    \label{fig:sntd_corner}
\end{figure*}

\subsection{Updated Time Delays and Magnifications from Template \hst Photometry}
\label{sub:sn_results}
We use the updated photometry from Table \ref{tab:im_mags} to constrain the time delays and magnifications for the multiple images of \lensedsn in the manner of \citet{rodney_gravitationally_2021}; \citetalias{pierel_lenswatch}. As summarized in \citetalias{pierel_lenswatch}, measuring the difference in time of peak brightness for each image directly \citep[e.g.,][]{rodney_sn_2016} is not possible with a single epoch, so we instead constrain the age of each SN image given a single light curve model. The relative age difference for each image is also a measure of the time delay, though we note this method is only possible because we have a reliable model for the light (and color) curve evolution as \lensedsn is a SN Ia.

We fit the photometry of the multiple images simultaneously using the \sntd software package \citep{pierel_turning_2019}, where we also include Milky Way dust extinction ($E(B-V)=0.16$ mag, $\ R_V=3.1$) based on the maps of \citet{schlafly_measuring_2011} and extinction curve of \citet{fitzpatrick_correcting_1999}. We also include additional uncertainty introduced by chromatic microlensing in the same manner as \citetalias{pierel_lenswatch}, which used the simulations of \citet{goldstein_precise_2018}. These are $\sim0.05, \ 0.05$, and $0.11\rm\,{mag}$ of additional color uncertainty in rest-frame $U-B$ ($\sim$F475W$-$F625W), $B-V$ ($\sim$F625W$-$F814W), and $U-V$ ($\sim$F475W$-$F814W) respectively. We add these uncertainties in quadrature to the color uncertainties from our photometric measurements for the fitting process. We note that this is by far the largest source of uncertainty in our time-delay measurements. We do not have estimates for chromatic microlensing uncertainty for the colors containing the F160W band; therefore, we fit only with the WFC3/UVIS bands, which is also a more direct comparison to \citetalias{pierel_lenswatch}. However, in Section~\ref{sub:f160w}, we will look at the predictions that an extended SALT2 model fit with the WFC3/UVIS bands makes for the F160W point and what this may tell us about chromatic microlensing.

\begin{figure}
    \centering
    \includegraphics[scale=0.6]{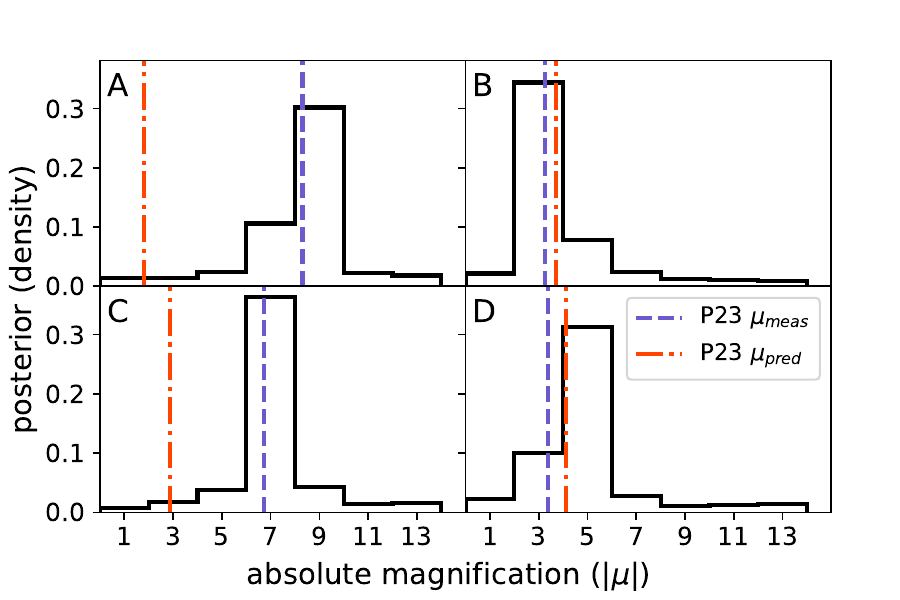}
    \caption{Posterior distributions of the absolute magnifications derived assuming a fiducial SN Ia standardization in black, obtained after fixing light curve and time delay parameters from Figure \ref{fig:sntd_corner}. The predicted magnifications from the joint lens model result and the measured magnifications from \citetalias{pierel_lenswatch} are shown as the overplotted dashed lines in red and blue respectively.}
    \label{fig:magnification_posteriors}
\end{figure}

We follow the methods outlined in detail by \citetalias{pierel_lenswatch}, originally adapted from the analysis of \citet{rodney_gravitationally_2021}, to measure time delays for \lensedsn. This process uses the SN color curves to constrain the time delay (with the \sntd ``Color'' method), and then fits for relative magnifications using a nested sampling method using \textsc{SNCosmo} and assuming a SALT2 SN model \citep{guy_salt2_2007,barbary_sncosmosncosmo_2016}. In this case, we will use the most up-to-date version in the literature \citep{taylor_revised_2021}. A well-sampled, unresolved light curve exists for \lensedsn and in \citetalias{goobar_zwicky} was fit with SALT2 to give $t_{pk}=\snpeak, \ c=\snc,$ and $ \ x_1=\snstretch$. We therefore follow \citetalias{pierel_lenswatch} and allow the $t_{pk}$ parameter, which here describes the time of peak for image A, to vary only within fifteen days of \snpeak. We also fix $x_1$ to the parameter derived by \citetalias{goobar_zwicky} ($x_1=\snstretch$), mainly to ensure an accurate light curve standardization. After these time delays have been measured with the \sntd Color method, we fix all best-fit parameters and find each image's apparent magnitude parameter, $m_B$, from a fit to the SALT2 model.

As a check, we also fit for time delays using \gaussn, a Gaussian Process-based time delay fitting code \citep{hayes_gaussn}. We use a version of \gaussn that allows us to leverage a template SED; in this case we use SALT2. To minimize systematics from template choice, GausSN allows for chromatic deviations from the SN light curve template. This added flexibility of the model leads to larger uncertainties on the time delay of SN Zwicky, as there is only one epoch of data per image to constrain any such deviations. Therefore, we have to rely on stronger assumptions about the true underlying shape of the light curve, as we do with \sntd, to get a more precise constraint on the time delay and magnification. Therefore, although we find consistent time delays with \gaussn as with the \sntd method, we proceed with our \sntd results for comparison with results from \citetalias{pierel_lenswatch}.

Following our above method, we convert the $m_B$ parameter values we obtained to absolute magnitudes by applying a fiducial SN Ia standardization to \lensedsn. Specifically, we apply light curve corrections using the SALT2 parameterization for stretch ($x_1=\snstretch$, with luminosity coefficient $\alpha=0.14$) and our measured color ($c = 0.11$ with a luminosity coefficient of $\beta=3.1$) in the manner of \citet{scolnic_complete_2018} to obtain absolute magnitude estimates. We then compare the distance modulus of each image to the value predicted by a flat $\Lambda$CDM model (with $H_0=70\,\rm{km}\, s^{-1}\, Mpc^{-1}, \ \Omega_m=0.3$) for an average SN\,Ia \citep[$M_B=-19.36$,][]{richardson_absolute_2014} at $z=\snz$ converted to the CMB frame. We combine the statistical uncertainties on each measured magnification with a systematic uncertainty based on the intrinsic scatter of SN\,Ia absolute magnitudes \citep[0.1\,mag;][]{scolnic_complete_2018}. Although different values for each of these coefficients could be used based on the study and sample, we choose these values to maintain consistency with \citetalias{pierel_lenswatch}. To be specific, we apply the standardization using the equation: 
\begin{equation}
\mu_\text{obs} = m_{B} + \alpha\,x_1 - \beta\,c - M_{B}
\end{equation}
where $\mu_\text{obs}$ represents the inferred distance modulus, and 
$\alpha$, $\beta$, and $M_{B}$ are the assumed values we discuss above.

The updated measured time delays and magnifications (with subscript ``updated'') are shown in Table \ref{tab:td_mu} compared with the measured and lens model-predicted values from \citetalias{pierel_lenswatch} (with subscripts ``P23 meas'' and ``P23 pred'' respectively). The model-predicted values we present are the joint modeling results from the ``Final" column of Table~3 in \citetalias{pierel_lenswatch}, which represents the weighted averages of the results from the individual model results, and are not from an individual lens model. The posterior distributions for all parameters fit with \sntd (using the conversions listed above) are shown in Figures \ref{fig:sntd_corner} and \ref{fig:magnification_posteriors}. While the relative time delays are too small and associated uncertainties too large to provide a useful direct cosmological constraint, these results provide a valuable check on our lens modeling predictions.

There are four lens models in \citetalias{pierel_lenswatch} that contributed to the final predicted value we discuss. Two of these models were constructed using \textsc{lenstronomy} \citep{birrer_gravitational_2015,birrer_lenstronomy_2018,birrer_lenstronomy_2021}, one with \textsc{lfit\_gui} \citep{shu_boss_2016}, and a final with the Gravitational Lens Efficiency Explorer (\textsc{glee}; \cite{Suyu_glee_2010,suyu_disentangling_2012,ertl_tdcosmo_2022}). All of these models relied on the image positions of \lensedsn to estimate the lens modeling parameters, as this system lacks gravitationally lensed arcs that could help constrain the models. The fluxes and standardizable nature of \lensedsn was used as a prior for a fifth lens model using \textsc{lenstronomy}, which is called the ``SALT+LS'' model. We will often discuss the combined or ``Final'' model results from the first four models, but we will also discuss this last model and its implications for our conclusions.

\begin{figure}
    \centering
    \includegraphics[scale=0.55]{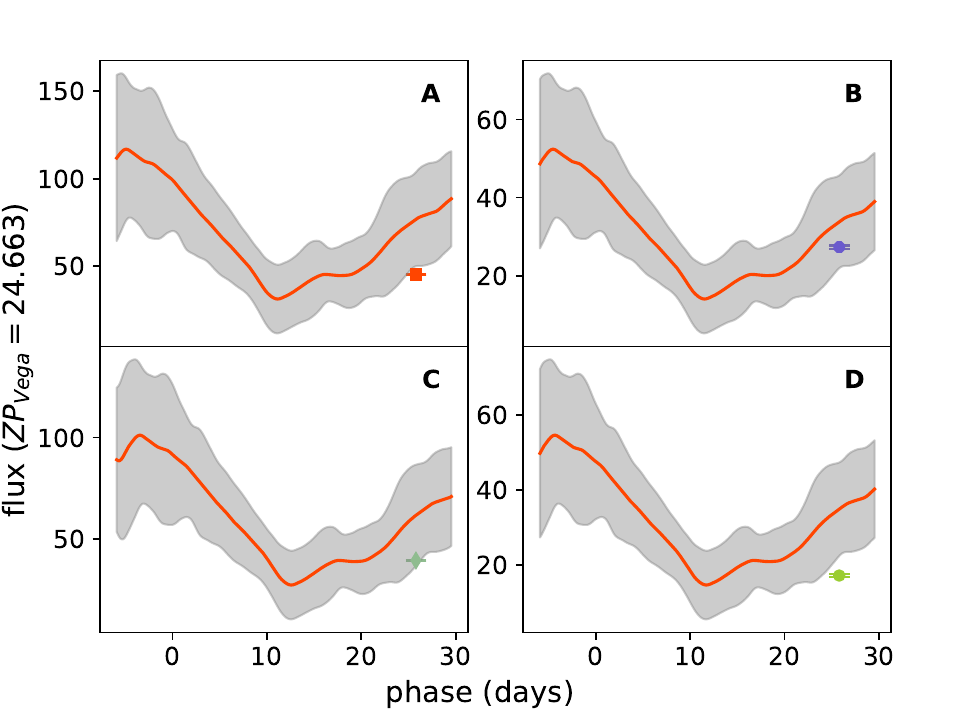}
    \caption{SALT2-extended fits to WFC3/UVIS filters, extrapolated to the WFC3/IR F160W filter as well as photometric measurements in the F160W filter as points for each of the four images. The solid red lines are the SALT2-extended model flux, while the gray regions show the model uncertainty.}
    \label{fig:salt2-ir}
\end{figure}

\subsection{Insights From F160W Photometry} \label{sub:f160w}

Microlensing simulations in the F160W band are beyond the scope of this project. Moustakas et al.~(in prep.) and Arendse et al.~(in prep.) will present in-depth microlensing analyses of \lensedsn. Nevertheless, with this new data we can now explore deviations from our lens model and SN light curve predictions in the near-infrared, which was previously not possible. 

We can take the value we obtain for color, $c=0.11$ from the \sntd method and create a prediction for the F160W band for all four of our SN images, fitting with the SALT2-extended model, a model that has trained SALT2 on SN Ia near-infrared (NIR) photometry \citep{kessler_snana}. We show the results of this method in Figure~\ref{fig:salt2-ir}. The SALT fits show that the predicted fluxes for all images appear to be higher than what we actually measure for our single epoch. However, there is a large amount of model uncertainty (shown as the gray regions in Figure~\ref{fig:salt2-ir}). For images A, C, and D, there is a $\gtrsim1\sigma$ discrepancy; however, for image B the offset is only about $0.5\sigma$. We also find a similar disagreement when fitting SALT3-NIR, another SALT template fitter that includes NIR training \citep{pierel_salt3nir}. Unfortunately, this model also includes significant model uncertainty at the epoch of our observations, thus providing a similar decrement in flux as with SALT2-extended.

\begin{figure}
    \centering
    \includegraphics[scale=0.55]{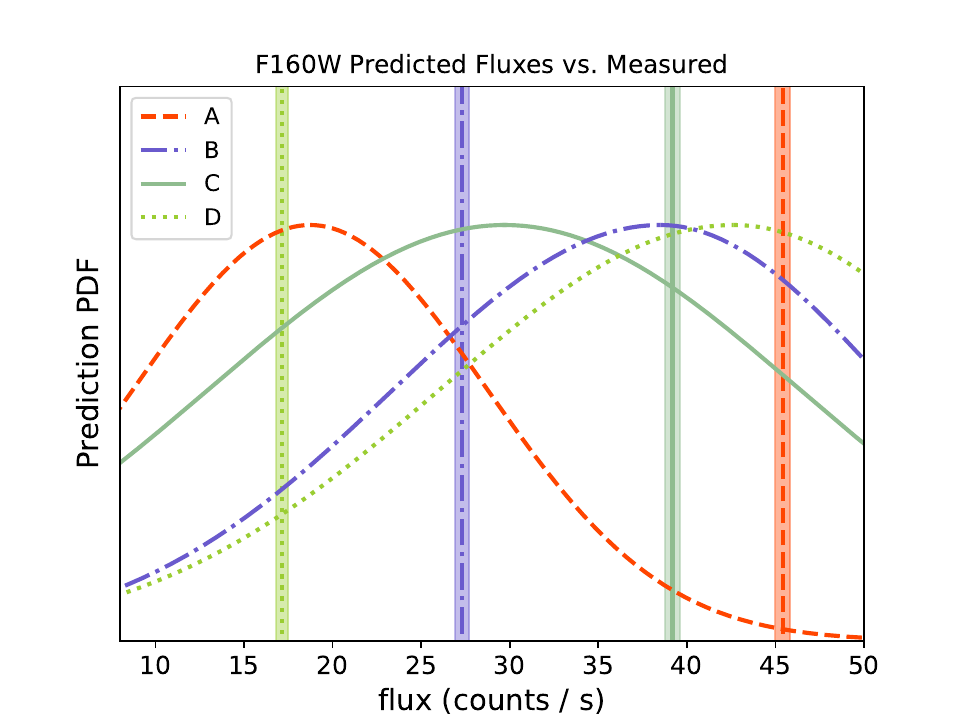}
    \caption{Measured WFC3/IR F160W fluxes from \dolphot with associated $1\sigma$ uncertainties as regions, with the joint model predicted fluxes for each image from \citetalias{pierel_lenswatch}, shown as the predicted probability distribution function for the flux of each image. Images B and C are consistent with lens model predictions to within $1\sigma$, while image A has the largest discrepancy.}
    \label{fig:mag_pred}
\end{figure}

Apart from our SN models, we can also look at the lens model predictions for individual image fluxes based on the total flux of the combined four images in the F160W band. Because we know that our time delays are small, on the order of $\sim1$ day, we approximate that the de-magnified flux of each image is the same. Therefore, we use the total integrated flux of all four images combined to predict the flux of each image using lens model predicted magnifications. For a total flux, $f_{tot}$, a demagnified flux that is equal for all images, $f$, and predicted magnifications for each image, $\mu_i: i \in \{ A | B | C | D \}$, the flux, $f$, and individual predicted fluxes with magnifications can be approximated as,

\begin{gather}
    f = \frac{f_{tot}}{\sum\limits_{i} \mu_i} \\
    f_i = f\,\mu_i
\end{gather}

\begin{figure*}
    \centering
    \includegraphics[scale=0.5]{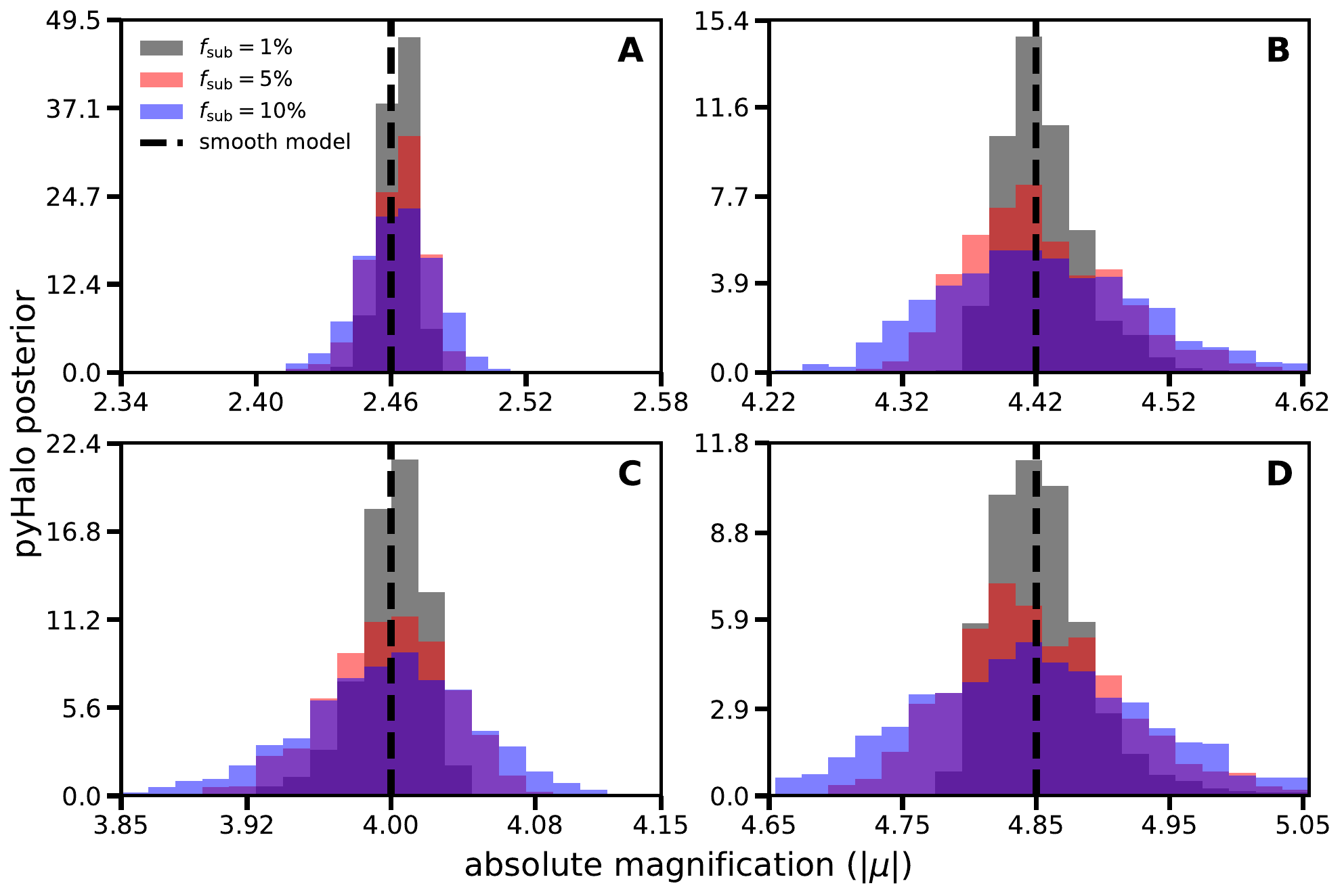}
    \caption{Posteriors of magnification estimate results from 1,000 realizations of \textsc{pyHalo} millilensing with dark matter subhalo fractions, $f_{sub}$, of $1\%$, $5\%$, and $10\%$ for the ``LS1" model from \citetalias{pierel_lenswatch}. We also show the non-perturbed results from a smooth lensing profile with \textsc{lenstronomy}. This shows that the uncertainty in magnifications for each image due to millilensing is quite small, on the order of a percent or less.}
    \label{fig:pyHalo_results}
\end{figure*}

We measure the total aperture flux of the four combined images in the F160W frames and use the predicted magnifications for each image from the joint lensing predictions in \citetalias{pierel_lenswatch} to make predictions for the fluxes of each image. We then compare these predictions to the measured fluxes from our \dolphot photometry. We show the results in Figure~\ref{fig:mag_pred}. We find that there is an agreement/disagreement between the joint modeling predictions and the measured photometry of $\sim 2.67\sigma,\, -0.74 \sigma,\, 0.57\sigma,\,$ and $-1.54\sigma$ for images A--D respectively, showing tight agreement for images B and C but a potential offset for images A and D. Each individual lens model also fails to predict the F160W flux for images A and D, while the joint predictions have the least significant offset. If we extend this analysis to the SALT+LS model as well, we find that the flux of image A is correctly predicted to within $1\sigma$, but there is still a significant offset for image D.

\subsection{Millilensing Test}

In order to help explain the discrepancies in absolute magnifications, we explore the possible presence of millilensing, where dark matter subhalos along the line of sight (LOS) to the images may be adding additional magnification to our flux estimates \citep{mao_substructure_1998,xu_substructure_2009,xu_substructure_2012}.

We use \textsc{pyHalo}\footnote{\url{https://github.com/dangilman/pyHalo}} \citep{gilman_pyhalo}, an open-source python package that generates realistic simulations of dark matter substructure around image positions and along the lines of sight to the SN images and measures magnifications using \textsc{lenstronomy}. Working off a framework made for the cluster analysis of \citet{pierel_h0pe}, we run three sets of 1,000 realizations to test the overall effect of millilensing. This model assumes a convergence and shear at each image position, which we set based on the results from the four main lens models in \citetalias{pierel_lenswatch} to test the impact of millilensing on each of them. We also set the bounds of simulated halo and subhalo masses to $m_{L} = 10^{5.5} M_{\odot}$ and $m_{H} = 10^9 M_{\odot}$, where $m_{L}$ is the lower mass limit and $m_{H}$ the higher. These limits were based on results from \citet{gilman_tdcosmo_2020}, who carried out tests with a lower minimum mass threshold and showed that any mass lower than that which we assume would not impact the results. Our upper limit is based on the assumption that halos more massive than $10^9 M_{\odot}$ would host a visible galaxy. The amount of substructure that is present in a \textsc{pyHalo} simulation is based on the parameter, $f_{sub}$, which sets what percentage of the dark matter that is impacting the image magnifications is in dark matter subhalos. To test three degrees of millilensing in the system, we assume three different values of $f_{sub}$: 1\%, 5\%, and 10\% for our three runs, which are based on both theoretical and observational results \citep{dalal_substructure_2002,gilman_pyhalo}. The simulated halos and subhalos have a Navarro-Frank-White (NFW: \citet{navarro_nfw_1996}) mass profile. For the lens plane subhalos, we assume a power-law subhalo mass function and place subhalos uniformly around each image. For the LOS halos, we assume a Sheth-Tormen mass function and place subhalos in a cylindrical volume around each image \citep{sheth_profile_1999}. The results of the simulations for the ``LS1" model from \citetalias{pierel_lenswatch} are shown in Figure~\ref{fig:pyHalo_results}, which presents the absolute magnification distributions from the three configurations of $f_{sub}$, as well as the smooth model results from \textsc{lenstronomy}. The uncertainties in the magnifications due to millilensing are of similar magnitude across all four models. The largest impact appears to be for image D, which shows a $1\sigma$ uncertainty that is a few percent of the median value. For images A and C, the images that have the largest discrepancies between predicted and measured magnifications across all lens models, there are uncertainties due to millilensing of $\lesssim 1$\% in the runs with the largest spread, $f_{sub}$ = 10\%. Thus, we find that the effects due to millilensing are negligible compared to the macromagnification uncertainties.

We can make sense of this small effect in the case of \lensedsn, as the redshift of the lensed SN is at \snz. This, combined with the fact that this is a galaxy-scale lensing system with a relatively small Einstein radius ($\theta_E \sim 0.168$": \citetalias{goobar_zwicky,pierel_lenswatch}), makes a chance alignment of a massive subhalo along the LOS less likely. There is a chance, however, that a dark matter subhalo could fall within one Einstein radius of a SN image, therefore causing a substantial difference in magnification, although our simulations show that this would be very unlikely. Such anomalous magnifications have been used to identify millilenses in the past \citep{diego_mothra_2023}. In the case of \lensedsn, this would have to account for the aberrant magnifications in both images A and C, which both lie close to the center of the lens galaxy, where tidal stripping and heating reduce the possibility of substructures that could act as millilenses.

\section{Discussion and Conclusions}
\label{sec:conclusion}
We have presented updated photometry in the \hst WFC3/UVIS F475W, F625W, and F814W bands for \lensedsn, as well as new WFC3/IR F160W photometry, by using template images obtained by the LensWatch collaboration. We use the resulting colors with an additional uncertainty due to chromatic microlensing to infer time delays of ($\redpen{0.52^{+2.11}_{-1.55}}$, $\redpen{1.97^{+2.28}_{-1.50}}$, $\redpen{0.78^{+2.03}_{-1.69}}$ days for images B-D relative to image A). \redpen{The time delays for images B and D are consistent with both the lens modeling predictions and measured values from \citetalias{pierel_lenswatch} to within <$\,\color{red}\redpen{1}\boldsymbol{\sigma}$. The time delay for image C is slightly elevated compared to \citetalias{pierel_lenswatch}, but is still consistent with the previously-measured time delay within $\color{red}\redpen{1}\boldsymbol{\sigma}$ and with the model prediction to within $\color{red}\redpen{2}\boldsymbol{\sigma}$}. Overall, the time delays are still on the order of one day or less. Using the color parameter, $c$ and time delays obtained with the \sntd color method, we fit the light curves for their apparent magnitudes and apply a fiducial light curve standardization to obtain absolute magnifications of $\redpen{9.13^{+5.21}_{-0.85}}$, $\redpen{4.04^{+1.85}_{-0.50}}$, $\redpen{7.79^{+3.88}_{-0.81}}$, and $\redpen{4.22^{+1.99}_{-0.46}}$ for images A, B, C, and D respectively. These values are within $1\sigma$ of the measured values from \citetalias{pierel_lenswatch}; however, they reinforce the tension between the formerly-measured absolute magnifications and the lens model predictions. \redpen{The absolute magnifications of images B and D are within $\color{red}\redpen{1}\boldsymbol{\sigma}$ of lens model predictions, but there is a statistically significant offset for images A and C of $\color{red}\redpen{5.91}\boldsymbol{\sigma}$ and $\color{red}\redpen{2.87}\boldsymbol{\sigma}$ respectively. Even for the SALT+LS model, which leverages the standardizability of \lensedsn as a prior on the lens model estimation, there is a $\sim3.5 \sigma$ offset between the predicted and measured magnifications for images A and C, even when including additional error due to milillensing. Thus, it is confirmed with the new template photometry that the magnification predictions of the lens models under-predict the measured magnifications of images A and C significantly.}

Before we examine other potential effects that could be adding to this measured discrepancy, we should also inspect what changes to the lens models themselves could close the observed gap in magnifications. For instance, because images A and C lie so close to the center of the galaxy, an added baryonic component of the lens model, with a more stretched light distribution than used in \citetalias{pierel_lenswatch} could impact the magnification measurements significantly \citep{diego_micro_2022}. For instance, if the lens galaxy were a disc galaxy, the disc could affect the lens model predicted magnifications \citep{hsueh_disk_2017}. We note, however, that we do not see any clear disc structure in the color image of the lens galaxy from the template images. This, along with the lens galaxy spectrum presented in \citetalias{goobar_zwicky} which shows only absorption features, makes a clear argument that the lens galaxy is elliptical.

Without an updated lens model to potentially explain the differences in magnification measurements and predictions, we must examine the effects of microlensing, millilensing, and/or differential dust extinction to explain our results \citep{metcalf_flux_2002,foxley-marrable_impact_2018,goldstein_precise_2018,hsueh_flux-ratio_2018,huber_strongly_2019}.

With updated template photometry, the evidence for differential dust extinction across the four images becomes more tenuous than in \citetalias{pierel_lenswatch}. Image C has the largest difference in F475W $-$ F814W color than the other images, with a difference of about $0.1\pm0.06$ mag, within $2\sigma$ of standard error. We also fit each SN image separately with the SALT2 model, fixing only the $x_1$ parameter and allowing the $c$ parameter to vary, which captures the combined effects of intrinsic color differences between SNe~Ia and the extinction from the intervening dust. We find $c$ values of $0.092 \pm 0.052$, $0.059 \pm 0.060$, $0.083 \pm 0.061$, and $0.067 \pm 0.061$ for images A-D respectively, which are consistent within $1\sigma$. We also fit each image with a SALT2 model that includes an additional extinction parameter of $E(B - V)$ from the lens galaxy. The $c$ values and $E(B - V)$ for these fits are all in very close agreement, well within $1\sigma$, with assumed $R_V$ values around 2.0. We then perform a final test by allowing $R_V$ to vary, and find close agreement among all three parameters: $c$, $E(B - V)$, and $R_V$. Based on these tests, there seems to be very little evidence for differential dust extinction affecting our photometry.

In order to test for millilensing as a possible explanation of the added magnification that we measure, we ran three sets of 1,000 realizations of a lensing simulation which included dark matter subhalos using \textsc{pyHalo}. \redpen{We find that the effect of the subhalos is negligible in the final magnification measurements, owing to the small cosmological volume over which the dark matter halos may affect our measurements. Therefore (unlike for a cluster-scale lensing system), we can assume that millilensing is not affecting our results at the scale of the discrepancy we are seeing. Therefore, we turn to microlensing as the remaining justification for the disagreement.}

With new photometric measurements in the F160W filter, we may potentially reveal interesting microlensing effects that could explain the discrepancy between lens model predictions and photometric measurements. As explored in Section~\ref{sub:f160w}, the F160W flux we measure for all images are lower than the prediction of the best fit SALT2 model to the optical data. For images A, C and D, this is a $\gtrsim1\sigma$ discrepancy, with image D being about $1.5\sigma$ lower than predicted. Image B is only about $0.5\sigma$ from the SN Ia model prediction. Due to the model uncertainty in the F160W wavelength regime at the phase of \lensedsn, we cannot conclusively say whether this evidence points to chromatic microlensing or not; however, this analysis hints that image D may be systematically less luminous than expected given the optical data and the SN Ia model. This cannot be due to de-magnification from chromatic microlensing, however, as the macrolensing magnification parity of image D is positive \citep{schechter_micro_2002}. In conjunction with the SN model analysis, we also compare these fluxes to what we would expect from the lens models presented in \citetalias{pierel_lenswatch} and find that images B and C are in good agreement with the lens models, to within $1\sigma$, while A is far above the predicted flux by $2.67\sigma$ and image D is below the predicted flux by $1.54\sigma$. We also note that images A and C are much closer to the center of the galaxy, thus increasing their chances of being microlensed by stars in the lens plane. Combining these two analyses, we can constrain a possible explanation for the predicted and measured magnifications, summarized below for each image:

\begin{itemize}
    \item \redpen{Image A:} The absolute magnification we measure for image A is $5.91\sigma$ higher than was predicted by the previous joint modeling analysis. The predicted flux in the F160W band based on the lens model predictions was also elevated by $2.67\sigma$. The measured F160W flux was slightly lower than expected given the SALT2 model fit, by about $1\sigma$. Therefore, there is strong support from the data that image A is experiencing significant microlensing, with a possible chromatic effect that is causing an excess of flux in the optical compared to the NIR.
    \item \redpen{Image B:} The absolute magnifications measured for image B, as well as the flux predictions for B from both the SN Ia model and lens model predictions are all within tight agreement. This indicates that image B is not affected by microlensing at any detectable level.
    \item \redpen{Image C:} There is a $2.87\sigma$ offset between the measured absolute magnification of image C and the lens model prediction from the optical filters. The F160W flux agrees strongly with joint lens model predictions, however. This, along with about a $1\sigma$ deficit in the SALT2 predicted flux, show that image C may be experiencing chromatic microlensing that is magnifying the optical bands much more than the NIR (similar to image A).
    \item \redpen{Image D:} Like image B, the absolute magnification measured for image D is in good agreement with lens model predictions. However, the predicted F160W flux from the lens models is $1.54\sigma$ lower than the predicted amount from the joint lens models, and there is a similarly significant decrement of the predicted F160W flux using the SALT model prediction. If the parity of image D was negative, this evidence could tentatively point toward the possibility of chromatic de-magnification. However, since the parity of image D is positive, this is not possible. Therefore, image D is likely not experiencing significant microlensing (similar to image B).
    
\end{itemize}

Our analysis in this paper shows that even for a strongly lensed SN Ia system with only one epoch of resolved photometry, such as \lensedsn, we are able to recover uncertainties on our time delays of only 1 - 2 days. Each individual image, as exemplified by this system, may be affected by microlensing differently, as the density and distribution of stars in the lens plane around the locations of the SN images is variable. However, if we are able to sample the light curve of a lensed SN earlier and with much more data, these effects will be greatly mitigated. 

The number of known galaxy-galaxy scaled strongly-lensed SN systems will increase by a few orders of magnitude with the upcoming Vera C. Rubin Observatory Legacy Survey of Space and Time \citep[LSST;][]{ivezic_lsst_2019} and \textit{Nancy Grace Roman Space Telescope} \citep{pierel_projected_2021} observations. This study shows that it will be imperative to maximize the number of systems we find that are in the $\sim3$ rest-frame week window of time post-explosion, where the effects of microlensing are ``achromatic'' \citep{goldstein_precise_2018}, thus limiting the type of systematic effects we wrestle with here (given that our observations are at $\gtrsim 6$ rest-frame weeks). And when such resolved, early-time observations are unavailable, follow-up campaigns from space such as those with LensWatch will be vital to creating the quality datasets we need to perform precise cosmological analyses in the future. A significant sample of these objects will allow for high precision measurements of cosmological parameters, including $H_0$. Thus, it is imperative to carefully study the sample of strongly lensed SNe that are currently available in preparation for such large-scale efforts.

\begin{center}
    \textbf{Acknowledgements}
\end{center}

This paper is based in part on observations with the NASA/ESA Hubble Space Telescope obtained from the Mikulski Archive for Space Telescopes at STScI. These observations are associated with program \#16264. CL acknowledges support from the National Science Foundation Graduate Research Fellowship under Grant No. DGE-2233066 and DOE award DE-SC0010008 to Rutgers University. JDRP is supported by NASA through a Einstein Fellowship grant No. HF2-51541.001 awarded by the Space Telescope Science Institute (STScI), which is operated by the Association of Universities for Research in Astronomy, Inc., for NASA, under contract NAS5-26555. This work has been enabled by support from the research project grant ‘Understanding the Dynamic Universe’ funded by the Knut and Alice Wallenberg Foundation under Dnr KAW 2018.0067. SD acknowledges support from a Kavli Fellowship and a JRF at Lucy Cavendish College. JMD acknowledges support from project PID2022-138896NB-C51 (MCIU/AEI/MINECO/FEDER, UE) Ministerio de Ciencia, Investigación y Universidade. CG is supported by a VILLUM FONDEN Young Investigator Grant (project number 25501). DG acknowledges support for this work provided by The Brinson Foundation through a Brinson Prize Fellowship grant. EEH is supported by a Gates Cambridge Scholarship (\#OPP1144). This work was supported by research grants (VIL16599, VIL54489) from VILLUM FONDEN. XH acknowledges the University of San Francisco Faculty Development Fund. This work is partly supported by the National Science Foundation of China (Grant No. 11821303 to SM). The work of LAM was carried out at the Jet Propulsion Laboratory, California Institute of Technology, under a contract with NASA. IPF acknowledges financial support from the Spanish Agencia Estatal de Investigaci\'{o} n del Ministerio de Cienciae Innovaci\'{o}n (AEI–MCINN) under grant PID2022-137779OB-C44. FP acknowledges support from the Spanish Ministerio de Ciencia, Innovación y Universidades (MICINN) under grant numbers PID2022-141915NB-C21. AJS was supported by NASA through the NASA Hubble Fellowship grant HST-HF2-51492 by the Space Telescope Science Institute, which is operated by the Association of Universities for Research in Astronomy, Inc., for NASA, under contract NAS5-26555. SHS thanks the Max Planck Society for support through the Max Planck Fellowship. This project has received funding from the European Research Council (ERC) under the European Union’s Horizon 2020 research and innovation programme (grant agreement No 771776). This research is supported in part by the Excellence Cluster ORIGINS which is funded by the Deutsche Forschungsgemeinschaft (DFG, German Research Foundation) under Germany’s Excellence Strategy -- EXC-2094 -- 390783311. TT acknowledges support by grant HST-GO-16264. This research made use of Montage. It is funded by the National Science Foundation under Grant Number ACI-1440620, and was previously funded by the National Aeronautics and Space Administration's Earth Science Technology Office, Computation Technologies Project, under Cooperative Agreement Number NCC5-626 between NASA and the California Institute of Technology.

\section*{Data Availability}

The data and scripts behind the figures in this study can be found at \url{https://github.com/Conor-Larison/lenswatch_ii} and in Zenodo at doi:\href{https://doi.org/10.5281/zenodo.14605501}{10.5281/zenodo.14605501}. The \hst data presented in this article were obtained from the Mikulski Archive for Space Telescopes (MAST) at the Space Telescope Science Institute. The specific observations analyzed can be accessed via \dataset[doi: 10.17909/s1n5-ym39]{https://doi.org/10.17909/s1n5-ym39}.

\software{\textsc{Astroalign} \citep{astroalign}, \textsc{Astropy} \citep{astropy_i,astropy_ii,astropy_iii}, \textsc{Astroquery} \citep{astroquery}, \textsc{Corner} \citep{corner}, \textsc{Colossus} \citep{colossus}, \textsc{DrizzlePac} \citep{drizzlepac}, \textsc{Dynesty} \citep{dynesty}, \textsc{IPython} \citep{ipython}, \textsc{Jupyter} \citep{jupyter}, \textsc{lenstronomy} \citep{birrer_lenstronomy_2018,birrer_lenstronomy_2022}, \textsc{Matplotlib} \citep{hunter_matplotlib}, \textsc{NumPy} \citep{numpy}, \textsc{pandas} \citep{mckinney_scipy,reback_pandas}, \textsc{pyHalo} \citep{gilman_pyhalo}, \textsc{SciPy} \citep{scipy}, \textsc{space\_phot} \citep{pierel_space_phot_2024}}

\bibliographystyle{aasjournal}
\bibliography{lenswatch.bib}

\end{document}